\documentstyle[epsbox,aps]{revtex}

\begin{document}
\newcommand{\ket}[1]{|\,#1\,\rangle}
\newcommand{\bra}[1]{\langle\,#1\,|}
\newcommand{\braket}[2]{\langle\,#1\,|\,#2\,\rangle}
\newcommand{\bold}[1]{\mbox{\boldmath $#1$}}

\title{Remote State Preparation without Oblivious Conditions}
\author{A.~Hayashi, T.~Hashimoto and M.~Horibe}
\address{Department of Applied Physics\\
           Fukui University, Fukui 910, Japan}
\draft

\maketitle

\begin{abstract}
In quantum teleportation, neither Alice nor 
Bob acquires
any classical knowledge on teleported states. 
The teleportation protocol is said to be oblivious to both parties.
In remote state preparation (RSP) it is assumed that Alice
is given complete classical knowledge on the state that is to
be prepared by Bob.
Recently, Leung and Shor \cite{Leung02} showed that 
the same amount of classical information as that in teleportation
needs to be transmitted in any exact and deterministic RSP protocol 
that is oblivious to Bob.
We study similar RSP protocols, but not necessarily oblivious to Bob.
First it is shown that Bob's quantum operation can be safely assumed
to be a unitary transformation.  
We then derive an equation that is a necessary and sufficient
condition for such a protocol to exist.  
By studying this equation, we show that one qubit RSP requires
2 cbits of classical communication,
which is the same amount as in teleportation, 
even if the protocol is not assumed oblivious to Bob.   
For higher dimensions, it is still open whether
the amount of classical communication can
be reduced by abandoning oblivious conditions.
\end{abstract}

\section{Introduction}
Interplay between classical information and quantum state 
shows non-trivial and remarkable aspects
when quantum entanglement is involved.
In quantum teleportation \cite{Bennett93},
one qubit in an unknown quantum state can be transmitted from a
sender (Alice) to a receiver (Bob) by a maximally entangled
quantum channel 
and two classical bit (cbit) communication. In order to teleport 
a quantum state in a $d$-dimensional space, $\log_2 d$ qubits,
Alice needs to transmit 
$2\log_2 d$ cbits of classical information to Bob. This is 
actually the minimum amount of classical communication, which
can be shown by combining teleportation protocol with another
striking scheme utilizing quantum entanglement,
superdense coding \cite{Bennett92}. 

In teleportation, neither Alice nor Bob acquires
any classical knowledge on teleported states. 
The teleportation protocol is said to be oblivious to Alice and Bob.
In remote state preparation (RSP), however, it is assumed that Alice
has complete classical knowledge on the state that is to be prepared
by Bob \cite{Lo00,Pati00,Bennett00,Zeng01}. 
The central concern has been whether quantum and 
classical resources can be reduced by Alice's knowledge on the 
state. 
In this respect, Lo has conjectured that RSP for a general state
requires at least as much as classical communication as
teleportation \cite{Lo00}.
An experimental implementation of RSP scheme has also
been reported \cite{Peng02}.

Recently, Leung and Shor \cite{Leung02} showed that 
the same amount of classical information as in teleportation
needs to be transmitted from Alice to Bob in any deterministic 
and exact RSP protocol that is oblivious to Bob.
Here the assumption that a protocol is oblivious to Bob means
specifically two things:
First, the probability that Alice sends a particular classical
message to Bob, does not depend on the state to be transmitted. 
Second, after Bob's quantum operation to restore the state,
the ancilla system contains no information on the prepared state.

In this paper we will study exact and deterministic RSP protocols
for a general state,
but not necessarily oblivious to Bob.
First we will show that Bob's quantum operation can be assumed to
be a unitary transformation.  
We then derive an equation that is a necessary and sufficient
condition for such a protocol to exist.  
By studying this equation, we show that 
in order to remotely prepare one qubit in a general state, 
Alice needs to transmit 2 cbits of classical information to Bob,
which is the same amount as in teleportation, 
even if the protocol is not assumed oblivious to Bob.   
For a general dimensional case, it is still open whether
the amount of classical communication can
be reduced by abandoning oblivious conditions.

\section{RSP protocol without oblivious conditions}
In this paper we only consider RSP protocols that are exact and 
deterministic. The diagram of protocol is depicted in Fig. 1.
\begin{figure}[htbp]
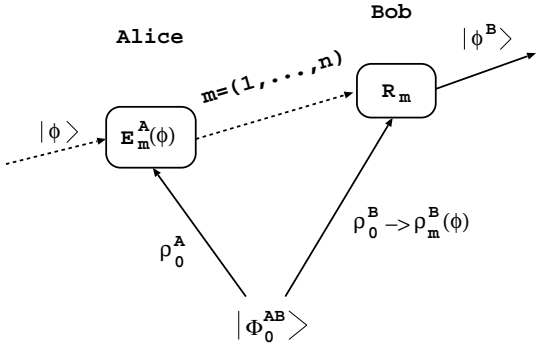

        \begin{center}
                \epsfile{file=fig1.eps,scale=0.6}
        \end{center}
        \caption{Remote state preparation diagram}
\end{figure}
The prior-entangled state shared by Alice and Bob is assumed 
to be a maximally entangled state in space AB defined by
\begin{eqnarray}
   \ket{\Phi_0^{\rm AB}} = \frac{1}{\sqrt{d}}\sum_{k=1}^{d}
              \ket{k^{\rm A}}\otimes\ket{k^{\rm B}},
\end{eqnarray}
where system A and B are $d$-dimensional Hilbert spaces,
with an orthonormal basis $\ket{k}\ (k=1,\ldots,d)$.
Writing $\rho_0^{\rm AB}=\ket{\Phi_0^{\rm AB}}\bra{\Phi_0^{\rm AB}}$
, we note that
   $\rho_0^{\rm A}\equiv {\rm tr_B}\rho_0^{\rm AB}= 
           \frac{{\bf 1}^{\rm A}}{d\ }$
and 
   $\rho_0^{\rm B}\equiv {\rm tr_A}\rho_0^{\rm AB}= 
           \frac{{\bf 1}^{\rm B}}{d\ }$.
Given a pure state $\ket{\phi}$ randomly chosen from an input
state space of $d$ dimension, Alice performs a POVM measurement
on system A with $n$ possible outcomes:
\begin{eqnarray}
    \sum_{m=1}^n E_m^{\rm A}(\phi)={\bf 1}^{\rm A}.
\end{eqnarray}
Remember that since Alice is assumed to have complete 
knowledge of state $\ket{\phi}$, the dependence of POVM elements
$E_m^A(\phi)$ on $\phi$ is not limited. The probability for Alice 
to obtain outcome $m$ is given by
\begin{eqnarray}
   p_m(\phi) = {\rm tr_A} \rho_0^{\rm A} E_m^{\rm A}(\phi).
   \label{eq_pm}
\end{eqnarray}
In this paper we do not assume the probability $p_m(\phi)$ 
is independent of $\phi$, implying the protocol may not be oblivious 
to Bob.
With outcome $m$, Bob's system B is given by
\begin{eqnarray}
   \rho_m^{\rm B}(\phi) = \frac{{\rm tr_A} \rho_0^{\rm AB} 
              E_m^{\rm A}(\phi)} {p_m(\phi)}.
   \label{eq_rhom}
\end{eqnarray}
Receiving a classical message $m$ ($m=1,\cdots,n$) from Alice,
Bob performs a trace-preserving quantum operation ${\cal R}_m$
on his subsystem B to restore the state $\ket{\phi}$:
\begin{eqnarray}
     {\cal R}_m(\rho_m^{\rm B}(\phi)) = 
                    \ket{\phi^{\rm B}} \bra{\phi^{\rm B}}.
\end{eqnarray} 

\section{It suffices for Bob to perform a unitary operation}
In this section we will show that Bob's quantum operation ${\cal R}_m$
is actually a unitary operation
${\cal R}_m(\rho) = u_m\rho\,u_m^+$, if the RSP protocol works 
for any state $\ket{\phi}$.  
First we observe the following theorem.

\bigskip
{\bf Theorem:}
Let ${\cal E}$ be a trace preserving quantum operation.
If for any state $\ket{\phi}$ there exists a density operator
$\rho_\phi$ such that 
    ${\cal E}(\rho_\phi) = \ket{\phi}\bra{\phi}$,
then the quantum operation ${\cal E}$ is a unitary operation
$    {\cal E}(\rho) = u\rho u^+$
, where $u$ is a unitary operator. 

\bigskip
Before proving the theorem, we note two general properties
of density operator, which will be used in the proof. 
The first one is that if ${\rm tr}(\rho\rho')=1$, then $\rho$
and $\rho'$ 
are identical and pure, which can be shown by the Cauchy-Schwarz
inequality.
Next, let $\rho^{\rm QR}$ be a density operator of a system consisting
of subsystems Q and R. Then the second property used in the proof
is that if
$\rho^{\rm Q}\equiv {\rm tr_R}\rho^{\rm QR}$ is pure, then
$\rho^{\rm QR} = \rho^{\rm Q}\otimes \rho^{\rm R}$, where 
$\rho^{\rm R}= {\rm tr_Q}\rho^{\rm QR}$.
This can be seen by observing subadditivity and the triangle 
inequality of von Neumann entropy $S$,
by which we find
\begin{eqnarray}
 S(\rho^{\rm R})&=&|S(\rho^{\rm R})-S(\rho^{\rm Q})| 
                         \le S(\rho^{\rm QR}),  \\
 S(\rho^{\rm QR})&\le& S(\rho^{\rm R})+S(\rho^{\rm Q})
                         =S(\rho^{\rm R}).
\end{eqnarray}
This means that equality in subadditivity  holds as
$S(\rho^{\rm QR}) = S(\rho^{\rm R})+S(\rho^{\rm Q})$, which is true 
only if $\rho^{\rm QR} = \rho^{\rm Q}\otimes \rho^{\rm R}$.

Now we will prove the theorem given in the above.

\bigskip
{\bf Proof:}
In the unitary model of a quantum operation,
the assumption in the theorem is stated as follows:
For any $\ket{\phi}$ there exists a density operator 
$\rho_\phi$ such that
\begin{eqnarray}
   {\rm tr_E}U
      \left(\rho_\phi \otimes \ket{0^{\rm E}}\bra{0^{\rm E}}\right)
        U^+
    = \ket{\phi}\bra{\phi},
\end{eqnarray}
where $\ket{0^{\rm E}}$ is a standard pure state of ancillary 
system E and $U$ is a unitary operator on the combined system.
As we have noted, if a subsystem is pure after tracing out the
ancilla system,
it is already pure in the combined system and therefore we have  
\begin{eqnarray}
   U
      \left(\rho_\phi \otimes \ket{0^{\rm E}}\bra{0^{\rm E}}\right)
   U^+
    = \ket{\phi}\bra{\phi}\otimes \rho_\phi^{\rm E}.
\end{eqnarray}
We will show that $\rho_\phi^{\rm E}$ is actually pure and 
independent of $\phi$.
Introducing an orthonormal basis $\ket{k}\ (k=1,\ldots,d)$,
we write
\begin{eqnarray}
   U
      \left(\rho_k \otimes \ket{0^{\rm E}}\bra{0^{\rm E}}\right)
   U^+
    = \ket{k}\bra{k}\otimes \rho_k^{\rm E}.
    \label{eq_k}
\end{eqnarray}
Multiplying the above Eq.(\ref{eq_k}) of index $k$ 
with the one of index $l$ 
and taking trace of the product, we find
\begin{eqnarray}
   {\rm tr}\,\rho_k\rho_l 
   &=& |\braket{k}{l}|^2 
           \,{\rm tr_E}\,\rho_k^{\rm E}\rho_l^{\rm E} 
                           \nonumber\\
   &=& \delta_{k,l}\,{\rm tr_E}\,\rho_k^{\rm E}\rho_k^{\rm E},
   \ \ (k,l=1,\ldots,d).
   \label{eq_kl}
\end{eqnarray}
This equation implies that the $d$ density operators 
$\rho_k$'s have orthogonal supports in the
$d$-dimensional space. This is possible only if 
$\rho_k = \ket{\psi_k}\bra{\psi_k}$, where the set
$\{\ket{\psi_k},\ k=1,\ldots,d\}$ is an orthonormal basis of 
the space.  We also find that $\rho_k^{\rm E}$ is pure, 
since  ${\rm tr}\rho_k^{\rm E}\rho_k^{\rm E}=1$.
    
In the same way as we obtained Eq.(\ref{eq_kl}), we find 
\begin{eqnarray}
   {\rm tr}\,\rho_\phi\rho_k 
   = |\braket{\phi}{k}|^2 
           \,{\rm tr_E}\,\rho_\phi^{\rm E}\rho_k^{\rm E},
   \ \ (k=1,\ldots,d).
\end{eqnarray}
Summing this equation over $k$ and
using $\sum_{k=1}^d \rho_k = 
\sum_{k=1}^d{\ket{\psi_k}\bra{\psi_k} =\bf 1}$, we obtain
\begin{eqnarray}
   1 = \sum_{k=1}^d |\braket{\phi}{k}|^2 
           \,{\rm tr_E}\,\rho_\phi^{\rm E}\rho_k^{\rm E},
\end{eqnarray}
which implies that $\rho_\phi^{E}$ is pure and given by 
\begin{eqnarray}
    \rho_\phi^{\rm E}= \sum_{k=1}^d |\braket{\phi}{k}|^2 
           \,\rho_k^{\rm E}.
\end{eqnarray}
From this we conclude that $\rho_k^{\rm E}$ is independent of $k$
and furthermore $\rho_\phi^{E}$ for a general $\ket{\phi}$ 
has no state dependence either.
Writing $\rho_\phi^{\rm E}=\ket{0^{'{\rm E}}}\bra{0^{'{\rm E}}}$,
we thus have
\begin{eqnarray}
 \rho_\phi \otimes \ket{0^{\rm E}}\bra{0^{\rm E}}
    = U^+\Big(
          \ket{\phi}\bra{\phi}\otimes 
          \ket{0^{'{\rm E}}}\bra{0^{'{\rm E}}}
      \Big)U.
\end{eqnarray}
Sandwiching this between $\bra{0^{\rm E}}$ and $\ket{0^{\rm E}}$
gives
\begin{eqnarray}
    \rho_\phi = u^+ \ket{\phi}\bra{\phi} u, \label{eq_uni}
\end{eqnarray}
where $u=\bra{0^{'{\rm E}}}U\ket{0^{\rm E}}$ and 
$u^+=\bra{0^{\rm E}}U^+\ket{0^{'{\rm E}}}$. It is clear that 
the operator $u$ must be a unitary operator 
since  $u^+ \ket{\phi}\bra{\phi} u$
is a density operator for any state $\ket{\phi}$.
Since Eq.(\ref{eq_uni}) holds for any $\ket{\phi}$, we conclude that
the quantum operation ${\cal R}$ is a unitary operation:
\begin{eqnarray}
     {\cal E}(\rho) = u \rho u^+. 
\end{eqnarray} 
\hfill $\Box$
\bigskip

Now remember that Bob receives classical message 
$m\in\{1,2,\ldots,n\}$ 
from Alice and performs a quantum operation ${\cal R}_m$ on the state
$\rho_m^{\rm B}(\phi)$ to restore the state $\ket{\phi}$
that Alice wants him to prepare:
\begin{eqnarray}
    {\cal R}_m(\rho_m^{\rm B}(\phi)) = 
           \ket{\phi^{\rm B}}\bra{\phi^{\rm B}}.
\end{eqnarray} 
Since this should hold for any state $\ket{\phi}$, by the 
theorem we have just proved, ${\cal R}_m$ turns out to be a
unitary operation:
\begin{eqnarray}
    {\cal R}_m(\rho) = u_m \rho\, u^+_m,
\end{eqnarray}
where $u_m$ is unitary.
We also note that we did not assume $\rho^{\rm E}$, the state of 
ancilla
system E after Bob's quantum operation, is independent 
of $\ket{\phi}$ (oblivious condition). But it was shown that 
$\rho^{\rm E}$ should be independent of $\ket{\phi}$ in the proof
of the theorem.

\section{Necessary and sufficient condition for RSP}
Now that we have shown that Bob's quantum operation is a 
unitary operation, we can derive an equation that is 
a necessary and sufficient condition for RSP protocols.

From Eq.(\ref{eq_pm}) and Eq.(\ref{eq_rhom}), we obtain
\begin{eqnarray}
   \sum_{m=1}^n p_m(\phi) \rho_m^{\rm B}(\phi) = \rho_0^{\rm B}
           = \frac{{\bf 1}^{\rm B}}{d\ },
\end{eqnarray}
which means that the density operator of system B should not change by 
Alice's POVM measurement on system A as long as an outcome of 
the measurement is unspecified. Using the result from the 
preceding section
$\rho_m^{\rm B}(\phi)=u_m^+ \ket{\phi^{\rm B}}\bra{\phi^{\rm B}}u_m$
, we get
\begin{eqnarray}
   \sum_{m=1}^n p_m(\phi)\,
       u_m^+ \ket{\phi^{\rm B}}\bra{\phi^{\rm B}}u_m
   =\frac{{\bf 1}^{\rm B}}{d\ }.
   \label{eq_RSPB}
\end{eqnarray}
Here $u_m$'s are unitary and $p_m(\phi)$ is the probability 
of outcome $m$ of Alice's POVM measurement, therefore 
$p_m(\phi) \ge 0$ and $\sum_{m=1}^n p_m(\phi)=1$.
And we note that this should hold for any state $\ket{\phi}$.

It is important that Eq.(\ref{eq_RSPB}) is also a sufficient
condition for RSP protocols.
Let us assume that Eq.(\ref{eq_RSPB}) holds in space B for some 
unitary operators $u_m$'s and for some probability distribution
$p_m(\phi)$, then the same equation
holds also in space A: 
\begin{eqnarray}
   \sum_{m=1}^n p_m(\phi)\,
       \ket{\phi_m^{\rm A}}\bra{\phi_m^{\rm A}}
   =\frac{{\bf 1}^{\rm A}}{d\ },
\end{eqnarray}
since the dimension is the same for spaces A and B,
where we wrote $\ket{\phi_m}=u_m^+\ket{\phi}$ for convenience.
Here for a state $\ket{\phi}=\sum_{k=1}^d\ket{k}\braket{k}{\phi}$,
we introduce 
the state $\ket{\bar\phi}$ defined as 
$\ket{\bar\phi}=\sum_{k=1}^d\ket{k}\braket{\phi}{k}$. 
Then it is clear that the following relation also holds:
\begin{eqnarray}
   \sum_{m=1}^n p_m(\phi)\,
       \ket{\bar\phi_m^{\rm A}}\bra{\bar\phi_m^{\rm A}}
   =\frac{{\bf 1}^{\rm A}}{d\ }.
\end{eqnarray}
From this relation we can construct POVM measurement elements as
\begin{eqnarray}
   E_m^{\rm A}(\phi) = d\,p_m(\phi)
        \ket{\bar\phi_m^{\rm A}}\bra{\bar\phi_m^{\rm A}}
   ,\ \ (m=1,\ldots,n).
    \label{eq_POVM}
\end{eqnarray}
Evidently each $E_m^{\rm A}(\phi)$ is a positive operator and 
$\sum_{m=1}^n E_m^{\rm A}(\phi) = {\bf 1}^{\rm A}$.
Since Alice is assumed to be given complete classical knowledge on
state $\ket{\phi}$, she can in principle implement
this POVM measurement.
The probability of an outcome $m$ is calculated as 
${\rm tr_A}\rho_0^{\rm A}E_m^{\rm A}(\phi)=
      \frac{1}{d}{\rm tr_A}E_m^{\rm A}(\phi)=p_m(\phi)$.
And with an outcome being given by $m$, the resultant state of B
is given by
\begin{eqnarray}
   \rho_m^{\rm B}(\phi) &=& d\,{\rm tr_A}
                      \ket{\Phi_0^{\rm AB}}\bra{\Phi_0^{\rm AB}}
   \cdot \ket{\bar\phi_m^{\rm A}}\bra{\bar\phi_m^{\rm A}} 
                                  \nonumber \\
                        &=&
       d\,\braket{\bar\phi_m^{\rm A}}{\Phi_0^{\rm AB}}
          \braket{\Phi_0^{\rm AB}}{\bar\phi_m^{\rm A}} 
                                  \nonumber \\
                        &=&
        \ket{\phi_m^{\rm B}}\bra{\phi_m^{\rm B}} = 
        u_m^+\ket{\phi^{\rm B}}\bra{\phi^{\rm B}}u_m.
\end{eqnarray}
Receiving classical message $m$ from Alice,  Bob can restore
the state $\ket{\phi}$ by a single unitary operation as
$u_m \rho_m^{\rm B}(\phi) u_m^+ 
= \ket{\phi^{\rm B}}\bra{\phi^{\rm B}}$. 

Thus Eq.(\ref{eq_RSPB}) is a necessary and sufficient condition 
for RSP protocols and will be called RSP equation hereafter
in this paper.

\section{RSP equation}
We will study the RSP equation (\ref{eq_RSPB}),
which is a necessary and sufficient
condition for RSP protocols:
\begin{eqnarray}
   \sum_{m=1}^n p_m(\phi)\,
       u_m^+ \ket{\phi}\bra{\phi}u_m
   =\frac{\bf 1}{d}.
   \label{eq_RSP}
\end{eqnarray}
Here superscripts A or B are omitted, since the equation should hold
in either $d$-dimensional space.

First we study the case that the probability $p_m(\phi)$ is independent
of $\ket{\phi}$, which is assumed in the paper by Leung and Shor 
\cite{Leung02}. We write the $(i,j)$-element of Eq.(\ref{eq_RSP})
explicitly:
\begin{eqnarray}
   \sum_{m=1}^n\sum_{k,l=1}^d
    p_m \bra{i}u_m^+\ket{k}c_kc_l^*\bra{l}u_m\ket{j}
   = \frac{1}{d}\delta_{ij},
             \nonumber \\
      (i,j=1,\ldots,d),
    \label{eq_explicitRSP}
\end{eqnarray}
where $c_k$'s are amplitudes of $\ket{\phi}$,  
$\ket{\phi}=\sum_{k=1}^nc_k\ket{k}$. 
It is convenient to introduce an
$n$ by $d^2$ matrix $X$ by 
\begin{eqnarray}
    X_{m;lj}&\equiv& \sqrt{dp_m}\bra{l}u_m\ket{j},
             \nonumber \\ 
   & &(m=1,\ldots,n),\ (l,j=1,\ldots,d),
\end{eqnarray}
and we can further rewrite Eq.(\ref{eq_explicitRSP}) as 
\begin{eqnarray}
     \sum_{k,l=1}^d c_k (X^+X)_{ki;lj}c_l^*
    = \delta_{ij}.
\end{eqnarray}
Remember that $c_k$'s are arbitrary apart from the normalization
condition and the matrix $X$ is assumed to be
independent of $c_k$. Therefore the matrix $X^+X$ must be a unit
matrix: $ (X^+X)_{ki;lj}=\delta_{kl}\delta_{ij}$.  
This implies the rank of $X$ is $d^2$ and consequently $n\ge d^2$.

Therefore the minimum amount of classical information, that Alice
needs to transmit to Bob, is at least $\log_2 d^2=2\log_2 d$
cbits in this 
oblivious case. This is the same amount of classical information
as the one in the teleportation. 
In the case of $n=d^2$, $X^+X={\bf 1}$ implies $XX^+={\bf 1}$
, from which we obtain
\begin{eqnarray}
    p_m= \frac{1}{d^2},\ \ \ {\rm tr}u_m^+u_{m'} = d\delta_{mm'}.
    \label{eq_trum}
\end{eqnarray} 
Therefore solutions are given by a set of unitary operators
that are complete and orthonormal with respect to 
the Hilbert-Schmidt inner product.  
As shown by Leung and Shor \cite{Leung02},
this gives a teleportation protocol, which is also oblivious to
Alice.
This is because Alice's POVM measurement Eq.(\ref{eq_POVM}) 
can be implemented 
as a state-independent projective measurement on a 
combined system of A and input space I: 
\begin{eqnarray}
   E_m^{\rm A}(\phi) &=& \frac{1}{d}\ket{\bar\phi_m^{\rm A}}
                                  \bra{\bar\phi_m^{\rm A}} \\
   &=& \bra{\phi^{\rm I}}u_m^{\rm I}
       \ket{\Phi_0^{\rm AI}}\bra{\Phi_0^{\rm AI}}
        u_m^{{\rm I}+}\ket{\phi^{\rm I}},
\end{eqnarray}
where
$\ket{\Phi_0^{\rm AI}} = \frac{1}{\sqrt{d}}\sum_{k=1}^{d}
              \ket{k^{\rm A}}\otimes\ket{k^{\rm I}}
$
and it is easy to verify that 
$u_m^{\rm I}
       \ket{\Phi_0^{\rm AI}}\bra{\Phi_0^{\rm AI}}
        u_m^{{\rm I}+},\ (m=1,\ldots,d^2)
$
are complete and orthogonal projectors.

One of the sets of $u_m$'s satisfying Eq.(\ref{eq_trum}) are 
shift operators in coordinate and momentum 
spaces \cite{Bennett93}:
\begin{eqnarray}
    u_m = u_{px}= e^{i\frac{2\pi}{d}p\hat x}
                  e^{-i\frac{2\pi}{d}x\hat p},
    \ \ (p,x=1,\ldots,d), 
\end{eqnarray}
where operators $\hat x$ and $\hat p$ are defined as 
\begin{eqnarray}
    \hat x = \sum_{k=1}^d k\ket{k}\bra{k}, \ \
    \hat p = \sum_{q=1}^d q\ket{\tilde q}\bra{\tilde q},
\end{eqnarray} 
and "momentum" eigenstates $\ket{\tilde q}$'s are given by
\begin{eqnarray}
    \ket{\tilde q} = \sum_{k=1}^d e^{i\frac{2\pi}{d}kq}\ket{k}.
\end{eqnarray}

Next we will study the case the probability $p_m(\phi)$ may depend 
on the state $\ket{\phi}$ that is to be remotely prepared. 
The question is whether this dependence can reduce the minimum
amount of classical communication.
In the case of one qubit RSP ($d=2$), we will show that this
is not the case:
the minimum amount of classical information turns out to be  
$2=2\log_2 d$ cbits as in teleportation. 

Unfortunately for general dimension $d$, however, we have only
limited results:
The RSP equation (\ref{eq_RSP}) immediately tells us that $n\ge d$,
which is known as Holevo's bound \cite{Holevo73},
since the equation requires 
that $\{u_m^+\ket{\phi},\ (m=1,\ldots,n)\}$ is complete in the 
$d$-dimensional space. We can also show that $n\ge d+1$. 
Suppose that the RSP equation (\ref{eq_RSP}) holds for $n=d$.
Generally in a $d$-dimensional space, the relation 
$\sum_{m=1}^d \ket{\chi_m}\bra{\chi_m} = {\bf 1}$
is satisfied if and only if the states 
$\ket{\chi_m}$'s are orthonormal. Therefore when $m\ne m'$,
the inner product 
$\bra{\phi}u_mu_{m'}^+\ket{\phi}$ should vanish for any 
$\ket{\phi}$, implying $u_mu_{m'}^+=0$. 
This, however, contradicts unitarity of $u_m$'s. 

Now we return to the qubit case ($d$=2). The Bloch sphere
representation is convenient for a pure qubit state 
$\ket{\phi}\bra{\phi}$: 
\begin{eqnarray}
   \ket{\phi}\bra{\phi} = 
      \frac{{\bf 1}+\bold{\chi}\cdot \bold{\sigma}}{2},
\end{eqnarray}
where $\bold{\chi}$ is a 3-dimensional unit vector and 
$\sigma_x,\sigma_y$ and $\sigma_z$
are the Pauli matrices. We also introduce a 3 by 3 rotation matrix 
$R_m$ for each unitary operator $u_m$ through 
\begin{eqnarray}
    u_m^+\sigma_i u_m = \sum_{j=1}^3 (R_m)_{ji}\sigma_j,
\end{eqnarray}
The RSP equation is then reduced to
\begin{eqnarray}
    \sum_{m=1}^n p_m(\bold{\chi})R_m \bold{\chi} = \bold{0},
      \label{eq_RSPn}
\end{eqnarray}
which should hold for any unit vector $\bold{\chi}$
and we emphasize again that the probability $p_m$ may depend
on $\bold{\chi}$.

It can be readily seen that
if Eq.(\ref{eq_RSPn}) holds for a set of rotation matrices $R_m$
and some probability $p_m(\bold{\chi})$, 
it is also satisfied by a set of 
transformed rotations $S R_m T$, 
with $S$ and $T$ being any rotation matrices,
and the probability $p_m(T\bold{\chi})$. 
With this freedom, we can safely assume that 
$R_1$ is a unit matrix, $R_2$ is a rotation about the $x$ axis, 
and $R_3$ is a rotation about an axis in the $xy$ plane.

Now suppose that Eq.(\ref{eq_RSPn}) holds for $n=3$ and take 
$\bold{\chi}=\bold{e}_x$ (the unit vector along the $x$ axis),
then we find  
\begin{eqnarray}
  \Big(p_1(\bold{e}_x)+p_2(\bold{e}_x)\Big)\bold{e}_x 
  + p_3(\bold{e}_x)\,R_3 \bold{e}_x = \bold{0}.
\end{eqnarray}
Since $p_m(\bold{e}_x)$ is a probability distribution,  
this equation is satisfied only when 
$R_3 \bold{e}_x = -\bold{e}_x$, namely 
$R_3$ is a rotation of $180^\circ$ about the $y$ axis.  
By a similar argument with $\bold{\chi}=\bold{e}_y$
(the unit vector along the $y$ axis), 
$R_2$ turns out to be a rotation of $180^\circ$
about the $x$ axis. Therefore, for a general unit vector
$\bold{\chi}=(\chi_x,\chi_y,\chi_z)$, Eq.(\ref{eq_RSPn})
with $n=3$ takes the 
following matrix form:
\begin{eqnarray}
   \left(\begin{array}{rrr}
       \chi_x  &  \chi_x  & -\chi_x \\
       \chi_y  & -\chi_y  &  \chi_y \\
       \chi_z  & -\chi_z  & -\chi_z \\
   \end{array}\right)
   \left(\begin{array}{c}
       p_1(\bold{\chi}) \\ 
       p_2(\bold{\chi}) \\ 
       p_3(\bold{\chi})  
   \end{array}\right)
  =
   \left(\begin{array}{c}
       0 \\ 
       0 \\ 
       0 
   \end{array}\right).
\end{eqnarray}
This equation has only a trivial solution 
$p_m(\bold{\chi})=0$ for $\bold{\chi}$ with 
$\chi_x \chi_y \chi_z \ne 0$, 
since the determinant of the matrix in the equation is
$4\chi_x \chi_y \chi_z$.
Thus we conclude that in order to remotely prepare a general qubit
state ($d=2$), Alice needs to transmit $2=\log_2 2^2$ cbits 
of classical information to Bob.

\section{Summary and Discussion}
In this paper we studied RSP schemes without assuming the 
protocol is oblivious to Bob. Bob's quantum operation was 
shown to be just a unitary operation, if the protocol works 
for a general state.  In this sense, Bob's operation is 
necessarily oblivious to himself. 

Using this fact we have derived the RSP equation, which is 
a necessary and sufficient condition for such an RSP protocol to
exist. 
By studying this equation, it was shown that 
in order to remotely prepare one qubit in a general state, 
Alice needs to transmit 2 cbits of classical information to Bob,
which is the same amount as in teleportation, 
even if the protocol is not assumed oblivious to Bob. 
So, for one-qubit RSP, Lo's conjecture \cite{Lo00} has been proved
without oblivious conditions. 

Unfortunately generalization to higher dimensions is not
straightforward. Though it is not yet clear 
whether the amount of classical communication can be 
reduced by abandoning oblivious conditions in higher dimensions. 
we believe that the RSP equation will be a key to obtain some
insights for further study in this direction. 

In this paper the input ensemble, from which the state
$\ket{\phi}$
is randomly chosen, is assumed to be the entire Hilbert
space of $d$ dimensions. 
We remark that
if the state is chosen from a sub-ensemble of the space,     
the RSP equation should still hold in the sub-ensemble,
as long as Bob's action can be assumed to be a unitary
operation.
In the case of qubits on the equatorial circle 
of the Bloch sphere \cite{Pati00,Bennett00},
the RSP equation (\ref{eq_RSPn})
with $n=2$ is satisfied as 
\begin{eqnarray}
    \frac{1}{2}(R_1\bold{\chi} +R_2\bold{\chi})={\bf 0},
\end{eqnarray} 
where $\bold{\chi}$ is a unit vector on the equator,
$R_1$ is a unit matrix, and $R_2$ is
a rotation of $180^\circ$ about the $z$ axis.
Generalizations of the equator and the polar great circle
to higher dimensions have been discussed by Zeng and Zhang
\cite{Zeng01}. We can also verify that corresponding RSP equations 
with $n=d$ are satisfied for those ensembles.

\end{document}